# GustosonicSense: Towards understanding the design of playful gustosonic eating experiences

### Yan Wang
Exertion Game Lab, Department of Human-Centred Computing, Monash University
Melbourne, Victoria, Australia
yan@exertiongameslab.org

### Humphrey O. Obie
Department of Computing Technologies, Swinburne University of Technology
HumaniSE Lab, Department of Software Systems and Cybersecurity, Monash University
Melbourne, Victoria, Australia
humphrey.obie@monash.edu

### Zhuying Li
School of Computer Science and Engineering, Southeast University
Nanjing, Jiangsu, China
zhuyingli@seu.edu.cn

### Flora D. Salim
School of Computer Science and Engineering, University of New South Wales (UNSW) Sydney
Sydney, New South Wales, Australia
flora.salim@unsw.edu.au

### John Grundy
HumaniSE Lab, Department of Software Systems and Cybersecurity, Monash University
Melbourne, Victoria, Australia
john.grundy@monash.edu

### Florian 'Floyd' Mueller
Exertion Games Lab, Department of Human-Centred Computing, Monash University
Melbourne, Victoria, Australia
floyd@exertiongameslab.org

## ABSTRACT

The pleasure that often comes with eating can be further enhanced with intelligent technology, as the field of human-food interaction suggests. However, knowledge on how to design such pleasure-supporting eating systems is limited. To begin filling this knowledge gap, we designed "GustosonicSense", a novel gustosonic eating system that utilizes wireless earbuds for sensing different eating and drinking actions with a machine learning algorithm and trigger playful sounds as a way to facilitate pleasurable eating experiences. We present the findings from our design and a study that revealed how we can support the "stimulation", "hedonism", and "reflexivity" for playful human-food interactions. Ultimately, with our work, we aim to support interaction designers in facilitating playful experiences with food.

## 1 INTRODUCTION

Pleasures associated with eating and drinking constitute some of life's most enjoyable experiences, said French gastronome Brillat-Savarin [18]. We were inspired by this, and also the fact that the HCI community has shown increased interest in Human-Food Interaction (HFI) [3, 84] and consequently explored ways in which intelligent systems can support dining experiences. For example, prior works designed applications for tracking people's food intake [56] in order to aid calorie counting as a way to manage diets [29] and to reduce food waste [40]. However, most of these prior works have focused on the technology or how to support instrumental goals (such as weight loss or sustainability), rather than considered how to support the eating experience itself. In response, a few research projects have recently emerged that proposed to also support the experiential aspects of dining, such as the pleasure and enjoyment that can come from eating and drinking [14, 19, 52]. To underline such a proposition, design researchers have developed, for example, a drinking cup that plays sounds when people drink out of a straw [87], seemingly supporting playful drinking experiences. This approach of using sound in combination with dining is based on the multisensory "gustosonic interaction" idea [82]. Building on this, we were inspired by Grimes et al.'s [27] article that asked for more "celebratory technology" as a way to support the experiential aspects of dining, in particular, playful aspects of eating. Playful eating is not just about energy and nutrient intake, but also a new way to celebrate meaningful everyday eating activities [5, 19]. Moreover, Mueller et al. [49, 52] proposed designing interactive technology to support eating as a form of play because people do not only eat to avoid starvation but also to feel food and pleasure [24]. As such, we draw from "gustosonic interactions" in combination with "celebratory technology" and respond to their respective calls for more design work: the result is "Gustosonic-Sense" (Figure 1). GustosonicSense is a novel gustosonic system that utilizes wireless earbuds for sensing different eating actions with a machine learning model and trigger playful sounds as a way to facilitate playful eating and drinking experiences.

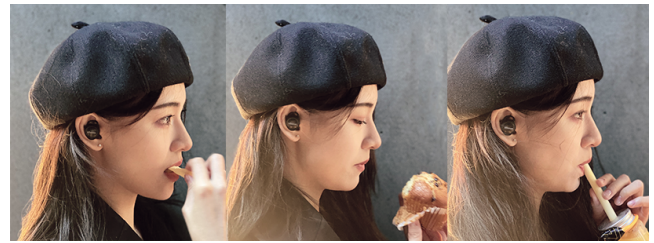

**Figure 1: The GustosonicSense system in use: we aimed for the system to be unobtrusive and allow the user to drink and eat anything they like.**

GustosonicSense allows the user to eat any types of food or drink any beverages, unlike prior work, which only supported specific foods [39, 85] or only supported drinking [32, 87]. Our system supports mobile use, while today's ubiquity of earbuds suggests that our system it can be used unobtrusively (unlike prior work, which required taping a jaw-movement sensor or microphone to the user's face [39]). These design features enabled us to conduct an in-the-wild study [62] where participants used GustosonicSense during their mealtimes. We gathered qualitative data from participants through semi-structured interviews to understand the user experience of our system. Based on an inductive thematic analysis [28], we were able to discuss the role "stimulation", "hedonism" and "reflexivity" played in facilitating playful eating and drinking experiences with our system.

Our work makes the following contributions:

- First, taking an exploratory analysis approach, we present GustosonicSense: a novel, mobile, unobtrusive gustosonic system that facilitates playful eating and drinking experiences. This system might inspire entertainment developers to utilize gustosonic interactions to facilitate playful experiences. This system might also inspire food designers to utilize sound to support existing dining experiences. Furthermore, this system might inspire headphone manufacturers to utilize their products' in-built sensors to facilitate augmented eating experiences. In future work, our system might also be appropriated by practitioners interested in changing eating behaviors, such as for healthy behavior change interventions.
- Second, we describe qualitative insights from our design process to assist designers and developers who are interested in creating future intelligent wearable systems for sensing eating and drinking actions, facilitating playful experiences, and delivering sound augmentations.
- Third, we articulate the results of our in-the-wild study to help researchers to better understand augmented playful experiences around eating and drinking by highlighting the joy that comes from multisensory experiences, which could be use for facilitating mindful engagement with what we eat and drink. This knowledge could be useful for interaction design and play researchers who are interested in understanding how to facilitate playful experiences through technology augmentations of everyday life practices.
- Fourth, we propose three design implications for the development of future gustosonic experiences to guide practitioners and designers interested in developing playful eating and drinking experiences.

## 2 RELATED WORK

In this section, we detail what we have learned from prior work that aimed to augment dining interactions, particularly through the use of sounds and playful eating/drinking interactions in HCI.

### 2.1 Augmenting dining interactions

Researchers have already begun to investigate how interactive technology could support dining experiences, particularly via sensing eating actions. For example, Zhang et al. [91] developed a 3D-printed pair of eyeglasses that embedded EMG electrodes, for sensing chewing actions and monitoring dietary adherence. Likewise, Bi et al. [15] designed an earpiece called "Auracle" that could recognize eating actions by capturing the sounds of a person chewing food. Similarly, Chun et al. [20] presented an augmented necklace that uses a multimodal sensing approach for detecting eating episodes. Moreover, Zhang et al. [92] designed "NeckSense", a multi-sensor necklace for detecting eating activities. The NeckSense system can detect chewing activity by measuring the distance between the device and the user's chin via a proximity sensor and detect feeding gestures via ambient light.

From these prior works, we learned that augmenting dining interactions with sensors can result in systems that provide value to eating and drinking experiences. However, the associated research focused primarily either on the technical implementation challenges or on instrumental aspects of eating, such as dietary monitoring. In contrast, we are interested in supporting the experiential aspects of dining. To support our investigation, we turned to prior work that appeared to consider the pleasurable aspects of eating and drinking, and these prior works seemed to often involve sounds, which we explain next.

### 2.2 Augmenting dining interactions with sounds

Recent research has found that people's enjoyment of food and beverages (and consequently their overall dining experience) is affected by what they hear while eating and drinking [71, 72, 75]. For example, the addition of sounds can influence people to perceive a food's texture as crunchy [78]. Sounds have also been used as an icebreaker to facilitate social eating experiences [74, 76]. Consequently, interaction design work has aimed to utilize sounds to enrich dining experiences. For example, Koizumi et al. [39] designed a system that requires attaching a sensor to the user's face to detect eating actions. If the sensor is triggered, the system plays a sound: in this case, a cartoon sound when the user chews a gummy sweet. More examples were presented by Wang et al. [85, 86]: "iS-cream" and "WeScream", comprises an interactive ice cream cone that randomly plays different sounds and musical notes when the user performs licking actions. Similarly, "Lickestra" is a musical art performance, which involves performers using licking actions (against crockery that was augmented with a capacitance sensor [11]) to improvise baselines and tones. Other examples include the "Drink Up Fountain" [42] (an augmented public drinking fountain that talks to people as they drink water from it) and a virtual reality experience [6], that uses chewing noises (detected by a sensor attached to the user's face) as a game controller. We learned from these prior works that sounds possess great potential to contribute to dining experiences. Furthermore, we are inspired by the associated designs of these prior works; they made us believe that sounds could be used as a valuable design resource when aiming to facilitate playful dining experiences. What we learned from prior HCI works concerned with such playful eating/drinking experiences, we discuss in the next section.

## 2.3 Playful eating/drinking interactions in HCI

We note that playful eating/drinking interactions can not only improve mental wellbeing, but increase happiness and support meaningful social interactions when people dine together [48, 85, 87]. Prior work argued that playful eating/drinking can bring joy and encourage people to engage all their senses [2, 23]. Building on this research suggested that combining eating/drinking interactions with digital technology could offer new ways to facilitate playful experiences that support "hedonism", "stimulation" ,and "enjoyment" [23, 89]. Hence, the play-focused HCI community and some art installations have contributed systems to investigate this [5, 19, 52, 89]. For example, Polotti et al. [60] designed a sonically augmented dining table that allows people to experience continuous sounds while having lunch. The dining table stimulates new ways of using cutlery and challenges the interaction with cutlery in an expressive manner. Another example is the work by Murer et al. [54]. The authors designed a playful lollipop system that can dynamically change the flavors of a lollipop, including flavors that are meant to surprise diners; these flavors are being used as a reward in a game. These works suggest that interactive technology can enrich playful eating/drinking experiences. However, although several works in the field of HFI have explored playful and pleasurable aspects of eating experiences [5, 14, 19], and the use of sounds appears to be a promising approach to enhance the experiential aspects of dining, previous works mostly focused on technical implementation details [39] and did not offer much insight into the associated user experiences. If prior work went beyond technical contributions, the associated systems mostly required dedicated setups with obtrusive sensor placements, making the capturing of relevant user experience data challenging [79]. Hence, our understanding of how to design for a playful multisensory eating experience is still underdeveloped. Without such knowledge, we believe that our ability to design less intrusive and more general-use systems is hindered. In order to address this knowledge gap, our work aims to make an initial contribution to our understanding of how to design unobtrusive systems that can facilitate playful eating and drinking experiences in everyday life.

## 3 GUSTOSONICSENSE

"GustosonicSense" is a novel gustosonic system that utilizes wireless earbuds to sense different eating actions and trigger different types of sounds: crunchy sounds (such as one might make when eating crisps), a series of notes from classical music, and fizzing sound commonly associated with carbonated water. The system is comprised of a pair of wireless earbuds [37] that feature a 6-axial inertial measurement unit (IMU) containing an 3-axial accelerometer and 3-axial gyroscope and a microphone. We used IMU data to sense temporalis muscle movements when the user engages in eating and drinking. A machine learning model recognizes the user's mouth activities including eating, drinking, speaking and idle activities. The earbuds communicate wirelessly with a custom-made mobile application. We developed the mobile application to analyze the data and stream the corresponding sounds back to the user while they eat and drink (Figure 2).

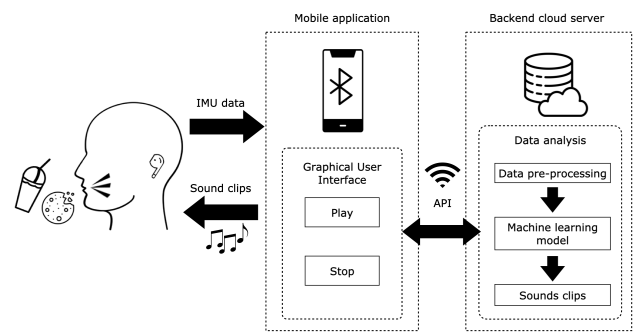

Figure 2: The system architecture of the GustosonicSense, showcasing the integration of earbuds and a cloud-based machine learning model.

## 3.1 Design consideration

To develop an unobtrusive gustosonic experience, we focused on the following design decisions, including the choice of detection technology, the choice of mouth activities and the choice of sounds.

*3.1.1 Choice of detection technology.* Inspired by prior work on eating detection [41], we utilize the earbuds to sense vibrations in the temporalis muscles when the user eats and drinks. This approach was designed to support the user in engaging in gustosonic interaction in their everyday life, without restriction on their dietary choices. The wearable nature of the earbuds enabled the use of the system at any time and in any place, allowing for real-time monitoring of eating actions in natural environments.

*3.1.2 Choice of mouth activities.* Prior research studies have investigated the influence of food textures on various parameters of the chewing process, such as jaw movement, facial and mouth muscle activity and eating rate [17, 81]. IMU signals from the earbuds can capture different levels of face and mouth movement when a user chews different types of food [34, 45]. In our system, we categorized eating and drinking activities into three distinct types based on food texture: hard food, soft food, and beverages. Additionally, we identified "speaking" as a fourth activity category. We did so because, in some eating situations, even when initially alone, people might engage in conversation with bystanders such as waiting staff or make phone calls. In a testing phase, we identified a fifth activity, "idleness," to demonstrate that the earbuds can be used in general everyday settings, such as listening to music or noise cancellation.

*3.1.3 Choice of sounds.* We choose three 60-seconds sound samples (crunchy sounds, classical notes, and carbonated water) intended to match three mouth activity categories: soft food, hard food, and drinking, respectively. This selection was motivated by prior work suggesting that designing incongruent sounds can elicit the user's curiosity while encouraging exploration and surprise, which, in turn, can facilitate playful eating and drinking experiences [83, 85, 88]. Therefore, we utilized crunchy sounds for the soft food activity and selected the classical notes for the hard food activity. We used an audio editor [8] to break each sound down into ten smaller 4-seconds clips and set up a random sequence within each activity to enhance the element of surprise, as surprise has been previously suggested as a way to facilitate playful experiences [7]. In our testing, we found that the users needed time to react to what they had heard. Consequently, after three rounds of iterations, we added a 1-second fade-out filter to each clip. The set of crunchy sounds consisted of crunchy chips sounds. The set of classical notes offered short, legato, and consonant piano chords. Finally, the carbonated water set offered bubbling and sparkling water sounds. When making these design decisions, we acknowledged that a user's understanding of sounds is influenced by their cultural conditioning and natural cognitive experiences mapping [9, 10], and we recognize that further investigation into sounds content might be needed.

## 3.2 Software: Data Acquisition, Analysis, Machine Learning Model and Application development

### 3.2.1 *Data acquisition.*
To detect the user's eating and drinking actions with the earbuds, we first created a database of four mouth activities from six participants (4 female, 2 male, all aged 24-33 years, none identified as non-binary or self-described). The database was used to analyze the diversity of mouth activities and frequency composition between users and throughout eating and drinking experiences. These activities comprised the eating of crunchy food, the eating of soft food, the drinking of a beverage, and speaking. We asked our participants to eat any food or any drink they liked for each activity. For example, a participant might eat chocolate or chips for the hard food eating actions, and yogurt or a banana for the soft food activity. Based on prior work relating to logging functionality using the earbuds [37], we first developed a logging application to collect the data. We also used data for the idleness activity from a related open-source project [33]. During the collection of the accelerometer and gyroscope data, the earbud's sampling rate was 50Hz. In total, there were 96,714 records across all five activities (Figure 3). Each participant's data collection time was 12 minutes, comprising three minutes for each of the four mouth activities. The received data was saved in a Comma-Separated Values (CSV) file.

### 3.2.2 *Data analysis and machine learning model.*
Using the data above, we trained and compared several machine learning models to identify an effective model for our system. We applied the cross-validation technique [90], which has been proven to be effective in addressing the risk of overfitting data. We trained the machine learning models and validated them using 10-fold cross validation. The result of the comparison of the F1 scores of the models is shown in Figure 4.

Based on the result of the comparison, the random forest classifier had the best performance on the dataset, with a mean F1 score of 0.85, followed by the Xgboost model, with an F1 score of 0.8. Next, using the randomized search CV, we tuned the hyperparameters of the best model (i.e., the random forest model) to get the best parameters and possibly increase the model's accuracy. However, the hyperparameter tuning increased the mean F1 score of the model from 0.85 to 0.86, only a slight improvement of 0.01. Table 2 shows the results of the random forest model.

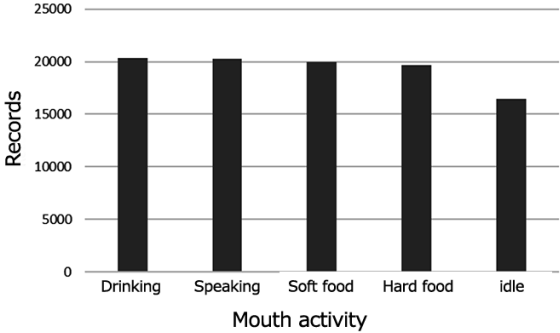

**Figure 3: Number of records for various mouth activities: drinking, speaking, consuming soft food, consuming hard food, and idle mouth activity.**

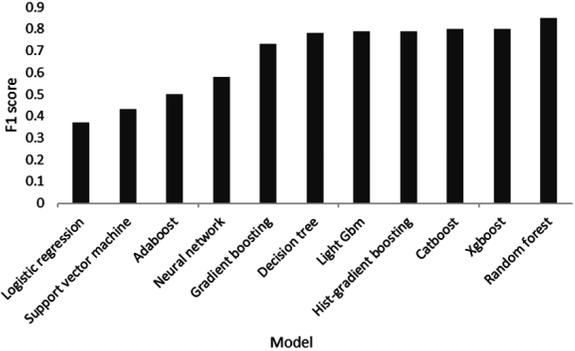

**Figure 4: Machine learning model comparison based on mean F1 scores using 10-fold cross validation.**

|  | Precision | Recall | F1 score |
|---|---|---|---|
| Crunchy food | 0.79 | 0.80 | 0.79 |
| Soft food | 0.82 | 0.81 | 0.82 |
| Beverages | 0.87 | 0.86 | 0.87 |
| Speaking | 0.83 | 0.83 | 0.83 |
| Idle | 0.99 | 1.00 | 1.00 |
| - | - | - | - |
| Macro average | 0.86 | 0.86 | 0.86 |
| Weighted average | 0.86 | 0.86 | 0.86 |

**Table 1: The results show that the mean F1 score from the random forest model is significant in our system.**

*3.2.3 Mobile application development.* We then developed a mobile application (app) that is capable of streaming data from and to the earbuds. To preserve battery life for our study, the machine learning model for detecting the eating activity was hosted on a backend server and accessed via a restful Application Programming Interface (API) service. As our research was conducted in an urban area with widespread 5G cellular coverage (and 4G as backup), we did not notice any significant delays as a result of using an external server. Once the earbuds were connected to the app and the app started by the participant, the IMU data was streamed to the app and then sent to the backend server in a JSON format, via an API call to detect the eating activity. Based on this data, the server made a prediction of the eating activity and returned a result. Using the eating activity prediction result, the app played the appropriate sound clip (or stream of sound clips) depending upon whether the result was the eating of hard food, soft food, or drinking (Figure 5). No sound was played if the system detected that the user was idle or speaking. The sound clips for each activity were not played sequentially, they were instead played randomly (within their activity) to foster unpredictability every time the system was in use. Because the average length of the sound clip is four seconds, the API call is made, the result interpreted, and the sound played every four seconds to avoid overloading the user with sounds.

that they could consume any food and drinks at any time and in any location they wished. The participants were encouraged to use our system as many times as possible. No compensation was provided. Table 2 provides participant details along with their professional and cultural background (no participants identified as non-binary or self-described). The cultural backgrounds of the participants are based on the places in which they grew up. The participants were all studied in the same city. No participants reported problems with hearing or any eating or drinking disorders. We received ethics approval from our institution to undertake this study.

| Name | Profession | Cultural background |
|------|-----------|---------------------|
| Rina (F, 25) | Design | Chinese |
| Karl (M, 24) | Accounting | Chinese |
| Bianca (F, 23) | Design | Singaporean |
| Lee (F, 26) | Art | Chinese |
| Lisa (F, 21) | Media | Australian |
| Gino (M, 29) | Information technology | Australian |

**Table 2: Participants in the in-the-wild study (pseudonyms used).**

We provided a package containing the earbuds, a smartphone and a portable router that supports wifi (Figure 6). We installed the app and paired the earbuds with the mobile phone before the study, so that participants only needed to tap a single button to start the system. We spent approximately 15 minutes explaining the study procedure to the participants and we also provided an instructional video that shows them how to set up and use the system in case they faced any technical challenges during the study period. We maintained contact with participants through emails and phone calls in case they needed technical support.

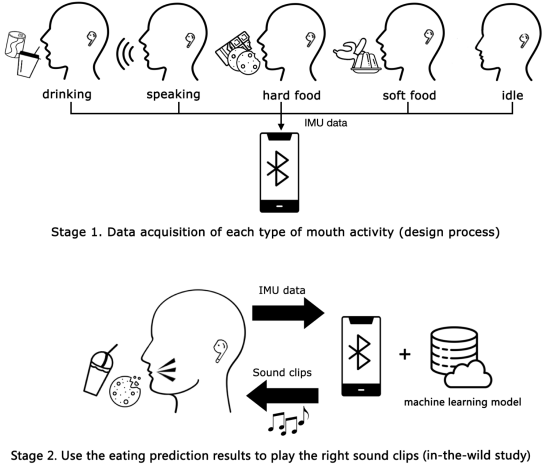

**Figure 5: The mobile application operates in two stages: Stage 1 involves data acquisition for each type of mouth activity; Stage 2 utilizes the eating prediction results to play the appropriate sound clips.**

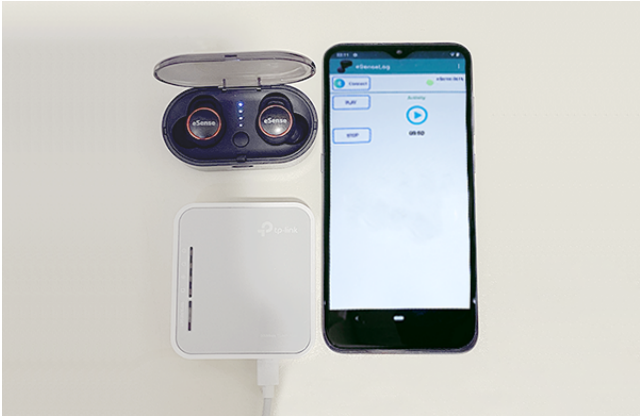

**Figure 6: Each participant received a package containing the GustosonicSense system that consists of the earbuds, the smartphone with our customized application, and a router.**

## 4 GUSTOSONICSENSE IN ACTION

To understand the user experience of GustosonicSense, we conducted an in-the-wild study [62] with 6 participants different to those identified above. Participant recruitment was undertaken using a combination of the snowball method [16] and convenience sampling. We told participants that they should use the system during at least three meals/snack times during a one-day study, and

# 5 DATA COLLECTION AND ANALYSIS

Our main sources of data were 30-minute interviews conducted with each participant at the end of the study. We used a semi-structured interview approach, leaving room for the discussion of other topics and supporting a deeper elucidation of participants' responses and thought processes [43]. During each interview, we took notes and recorded the audio, which was later transcribed. We asked questions related to the research aims to give focus, while leaving sufficient flexibility in the discussion for other matters to be raised. The questions focused on participants' motivations, expectations, and experiences of using GustosonicSense, including: how participants used the system; what kinds of food and beverages they consumed during the study; how the augmented sounds made them feel; and how the sounds affected their eating and drinking experiences; and what insights participants had regarding their eating and drinking experiences with the system and any interesting stories that arose from using those experiences. Additionally, we welcomed opportunities to view any photographs or recordings that participants might have captured while using GustosonicSense. This data helped us understand how the participants reacted to GustosonicSense while they ate and drank.

We used NVivo [57] to undertake an inductive thematic analysis [63]. We combined each question and each participant's answer and considered them to be one unit of qualitative data. In total, there were 183 units of data. Two researchers read the transcripts three times to become familiar with the data and then coded it independently. Their coding helped the team to identify and group the most interesting features of the data units. In the first round of the thematic analysis, we develop 12 code labels. In the second round, two researchers discussed and re-examined the codes to merge similar ones and reduce complexity. These labels were then iteratively clustered into three higher level themes with the help of one senior researcher.

# 6 RESULTS

Overall, the GustosonicSense system appeared to function as intended (further explained below) when the user consumed different types of food and drank different types of beverages, with the result being a playful eating and drinking experience. The results of the study suggest that the GustosonicSense system can facilitate a playful dining experience and that it achieved this outcome by fostering the "stimulation", "hedonism", and "reflexivity", each of which our analysis identified as themes.

## 6.1 Stimulation

In our study, all participants enjoyed the GustosonicSense experience and reported upon hearing the sounds excitement, which appeared to stimulate them in a positive manner.

### 6.1.1 Delayed sound to facilitate surprise.
The GustosonicSense system did not always play sound immediately when users started eating or drinking, which resulting in a delayed sound experience. This delay was not always considered a problem. For example, Rina liked that the sound came on sometime during the eating: *"Interesting, it was nice to experience the playful sounds sometime while eating."* She explained that sometimes the sound would only come on after having eaten a bit (possibly as a result of her initially not moving her temporalis muscle very much), and that, when it did, it was a welcome surprise that led to excitement.

### 6.1.2 Incorporating the earbuds into the body schema.
The choice of using earbuds appeared to contribute to the stimulation that participants appreciated, especially when we compare the earbuds to much larger and heavier traditional headphones. It seemed that the form factor and weight of the earbuds allowed some participants to incorporate them into their body schema [21, 47], sometimes to the extent that they forgot that they were wearing them. When the sounds were played because of eating or drinking, they surprised the participants. For example, Lee explained how he forgot that he was wearing the earbuds: *"When I ate as usual, I forgot the earbuds can playback the sounds. When the sounds suddenly came out, I felt surprise, but exciting"*. However, we also note that this experience was not the case for all participants. Four users reported that the earbuds were uncomfortable when worn for a long duration, their complaints included: *"I felt some weight of the device in my ear if I wore the earbud for a longer time"*, and: *"I forgot about it [earbuds] quickly at the beginning. Although I enjoyed those sounds surprising me, I felt ache around my ear canal after wearing it for half an hour."*

### 6.1.3 Sound change as a result of erroneous detection.
The choice of using different sound sources also appeared to contribute to participants' positive experiences. Participants reported that they did not consider it an error when the sounds changed (this could occur when the detection system erroneously identified two different food types even though it was the same food; and this change sometimes occurred when users chewed for rather long periods). On the contrary, four participants reported expressively that they enjoyed it if there was a sound change while they ate the same food item. Karl explained how this occurred when he was chewing pumpkin: *"When I was chewing pumpkin in my lunch, the sounds changed between the chips sound and the pleasant piano notes. I enjoyed the pleasant piano sounds."*

### 6.1.4 Non-matching sounds stimulating taste experience.
Participants also appreciated that the sounds appeared to enrich their eating experiences. Even though (or perhaps, because) the sounds did not match what participants expected of the food to sound like, the sounds still seemed to complement and hence enhance the dish they were eating. For example, Gino said: *"The crunchy sounds were funny and playful, and enhanced my taste of pasta nicely."* This result aligns with prior work, which already suggested that playing sounds at the same time as eating can enhance the taste experience [83, 87].

### 6.1.5 Supporting different meal scenarios.
Participants appreciated that they were able to experience the system in different meal scenarios. For example, users reported that they liked being able to use the system during regular lunch time, for example, when eating alone in a food court. *Bianca said: "I used the earbuds when eating alone in a canteen and food court. It could cancel the ambient noise when I was eating."* The system allowed participants to focus on the food (in contrast to, for example, engage in social media scrolling, which might distract from the food) and experience a more pleasant soundscape than offered by the food court noise. Furthermore, participants found value in using the system for less regular mealtimes, such as quick snacks, because the short sounds appeared to match the rapid nature of the activity.

## 6.2 Hedonism

All participants reported that they felt pleasure because of engaging with the system via eating and drinking. The user experiences appeared to align with the seeking of pleasure by participants for themselves while being self-indulgent via eating and drinking.

### 6.2.1 Supporting being self-indulgent when alone instead of sharing the food with others.
Participants appreciated that the system supported them to be self-indulgent with their food and alone. For example, Lisa told the story about the system supporting her enjoyment of a particular food she likes very much, which she was eating when alone and could not share with anybody: *"It was really cool. I liked those playful sounds while I ate out alone. I could use it in a public space, because it was a earbuds"*. Lee commented on the relationship between being self-indulgent and eating in relation to other people being present. While it appears easy to share the enjoyment of food when others are present, it is, by definition, not possible to do so when alone. Participants agreed that GustosonicSense would not be suitable as additional dining partner to share such enjoyment, however, it appeared to function as an in-between compromise: not as good as sharing enjoyment of a dish with another person, yet better than enjoying a dish alone. Lee summarized these views: *"I would not use it [the system] at home, because I shared meals with my partner. But when I ate out alone, I used it as a way to have pleasurable experiences."*

### 6.2.2 Supporting being self-indulgent without screens.
During the interviews, the topic of the system not requiring the use of visual elements often came up. Participants commented that they expected that the mobile phone application provides visual feedback or inform them of some data insights, because this is what they would expect of an app. It appeared that participants appreciated that there were no *"visual elements while using the system"* (Bianca), which seemed to support them being self-indulgent with their food, because all participants agreed that the use of screens distracted them from their food and drinks and limit any opportunities to gain pleasure through eating and drinking.

### 6.2.3 Wondering if sounds could affect eating and drinking perceptions.
It appeared that hearing sounds that do not resemble what a food is expected to sound like supported the pleasures that participants derived from eating and drinking. In particular, participants reported that hearing incongruent sounds when biting into food or sipping from drinks supported their food perceptions. This aspect of their experience appeared to be heightened when the system misidentified food items, such as those moments when drinking was recognized as eating crunchy food. For example, Bianca was one of two participants who really liked the crunchy sounds. She explained how the system made her perceive water as crunchy chips: *"The sounds were interesting [...] I enjoyed the carbonated sounds so much but sometimes it [the system] could detect wrong food [with laugh]. I expected to hear the carbonated bubbling sounds, but the system delivered the piano notes. I felt that the system might recognize my water was crunchy."*

## 6.3 Reflexivity

Participants reported that they appreciated that they could use the system in their own time and independently from the research setting, and that they were able to explore how they could creatively use the system to make it work for them.

### 6.3.1 Eating any type of food or drinks.
Our participants appreciated that the system allowed them to eat any types of food or drinks. Some participants appeared to be nervous during our initial conversations with them, when they learned that we are were investigating food interactions, and they felt compelled to mention that they either had dietary requirements (such as being a vegetarian or vegan), or that they had food allergies. Upon hearing that they would have complete control over what they ate and drank, they seemed relieved and much more relaxed for the remainder of the conversation.

### 6.3.2 Engaging with the system at anytime.
Our participants appreciated that they could use the system at any time, often making last-minute decisions whether to engage with it or not. For example, participants stopped using the system by taking out the earbuds when, for example, they wanted to talk to someone passing by. Participants liked that they did not have to press an additional button to stop the system, and that the termination of the sensing of their data (which could be perceived as personal data as so close to the body [44, 51, 53]) was triggered by the same action as taking off the earbuds. Participants reported that it was sometimes simply not appropriate to wear earbuds in some social settings, such as when visiting higher-end venues, even when alone, or when others might want to talk to you, even though people might sit on their own, such as in a workplace canteen. Our participants also reported that changing environmental conditions might demand that they take the earbuds off, for example, when a siren starts, or if waiting staff ask them a question. Furthermore, participants mentioned that hearing the sounds for longer might decrease the novelty that they perceived, and that they might find that they want to enjoy the rest of their meal without the system (possibly with a more heightened sense of appreciation of their dish). Overall, our participants appreciated the opportunity to fully control when they engaged with the system, and when they stopped.

### 6.3.3 Playing with sounds via eating and drinking.
The participants often used the word "play" when they described their interactions using the system. For example, Gino said: *"The system was definitely much more interesting than using normal earbuds. I could eat any food and beverages and play with those sounds while eating and drinking."* It appeared that participants considered using the system as a form of play, during which they controlled if, when, and how they chewed and sipped. All participants experimented with different ways in which they could consume their foods and drinks to explore what effects they could have on the system, such as the sounds they could produce. In this way, we contend that the system seemed to support participants' abilities to perform with creativity (by exploring creative ways to eat).

### 6.3.4 Sounds affected eating perception.
Our interviews revealed that participants not only explored different eating actions to trigger the sounds they would hear, but that the sounds they heard also affected their eating actions. The system appeared to support this effect by allowing people to eat and drink any type of food and in any way they wanted. This loop of "eating and drinking affecting sounds-sounds affecting eating and drinking" appeared to support independent acting, and we suggest that the playful character of the experience might speak nicely to people "performing with creativity", which is known to facilitate reflexivity [13]. For example, Lee explained that the sounds made them continue chewing: *"I felt that my strawberry pastry tasted like crisp chips when I heard the crunchy sounds. I could not stop chewing the pastry."* Similarly, Rina reported chewing more than usual as a result of hearing the sounds. This participant even reported a change in appetite: *"Although some chaotic sounds mixed when I ate or drank, I could see my appetite changed when engaging with the system. I tried to match each bite with the sounds; I chewed food more times than usual."* Karl reported learning about how sounds affect our eating as a result of having participated in the study, realizing that the food sounds could make him eat more slowly: *"I never recognized that food sound can affect our eating experience before this study: I could imagine how these sounds affect our eating experiences in some social places. To match the sound I heard, I felt I ate noodles slower than usual."*

## 7 DISCUSSION

We believe that the above results contribute to our understanding of the design of playful gustosonic experiences. We have shown through our analysis how specific design choices can facilitate playful experiences when users engage with an interactive eating system. We now discuss our results in relation to prior work.

### 7.1 Experiencing eating as play can offer more surprises

The GustosonicSense system offered opportunities for participants to intuitively experience how the earbuds delivered different sounds when consuming various types of food and beverages. Prior work has pointed out that designing technology around food should pay more attention to intuitive eating for pleasurable and meaningful eating and drinking experiences [2, 89]. Moreover, Mueller et al. proposed that designing digital technology around food can reframe eating and drinking activities as play to facilitate more engaging dining experiences [52]. In our study, we extended these prior works by leveraging a wearable system (the earbuds) with a machine learning algorithm to support playful eating experiences. Our system allows the participant to experience eating as play via the earbuds without requiring a specific food as an interface. Any types of food and beverages can be used as an interface to facilitate more exploration. In addition, the employed machine learning model enabled participants to eat and drink in various environment (in contrast to, for example, a more rigid rule-based system that would only support a particular food or a specific social setting [48]). While the prediction of the eating activities was not perfectly accurate, our implementation provided participants with an unobtrusive interaction, which seemed to contribute to an intuitive experience as play. Furthermore, although we did not play back a wide range of sounds, the resulting sounds feedback often surprised participants. They experimented with eating different types of food and beverages and adjusted their eating and drinking actions intuitively, such as changing chewing sequence, adjusting eating pace and muscular work. As such, our work demonstrated that even a system utilizing a technology as ubiquitous as earbuds has the potential to encourage people to explore new ways of experiencing food and drinks as play in everyday settings.

*7.1.1 Design implication: employ unobtrusive technology for meaningful eating.* We recommend designers to employ unobtrusive technology when designing interactive eating experiences for meaningful eating. Prior work proposed that unobtrusive sensing can enhance the comfort of using digital technology for continuous experiences [93] and increase the effectiveness of recognition processes in everyday life [80]. In our work, we utilized the earbuds that allowed for an intuitive eating experience associated with sounds. Participants were able to engage in gustosonic interactions that do not require prior knowledge or learning phase of the technical system, facilitating immediate surprises and taste explorations. The earbuds do not merely have the ability to recognize eating behaviors but also facilitate on-the-spot exploration and savoring with any food or drinks participants encountered. Therefore, we propose to designers to consider unobtrusive technology for exploratory eating, as it appeared it can enable experiencing eating as play.

### 7.2 Incongruence between food textures and sounds can elicit curiosity

People can perceive the food properties through different sensorial modalities because our brains can connect information from multisensory inputs, while the sensory information usually corresponds to the same identity [68, 83]. For example, people can hear the sound of sea waves and associate this with features of the sea, such as its blue color, seafood and saltiness [22]. This congruent experience occurs when our perception matches our expectation [68]. In contrast, feelings of incongruence can occur when people perceive mismatched sensory experiences. In the field of HCI, researchers have used the benefits of incongruence for enriched user experiences. Designing appropriate incongruence can evoke humor and curiosity to enhance hedonic experiences [64]. In our study, participants initially experienced a feeling of congruence between food textures and their expectations of food perception before engaging with our system. However, when participants encountered an incongruent sound that differs from what whey they had originally expected, they reported surprise, fun and enjoyment, resulting in a sense of curiosity about what caused this incongruence. The participants then often spontaneously began to discover new ways of interacting with food. They appreciated the opportunity to change their eating activity in the moment and reflect on their eating actions. They realized, through incongruent sounds, what food they had eaten, and they began to adjust their eating actions.

*7.2.1 Design implication: embrace incongruency as a design resource.* We recommend designers to embrace incongruency as a design resource for surprise and curiosity, facilitating playful eating experiences. Prior work proposed that mismatched auditory feedback with the flavor and texture of food can evoke a sense of wonder and playfulness [67, 77]. Our study suggested that this incongruent experiences can transform eating into an enjoyable activity facilitating a deeper interest in the food and its properties. Moreover, by creating an unexpected sensory experience, participants were encouraged to pay closer attention to their eating actions. This heightened awareness might lead to a better understanding of their food choices and potentially promote mindfulness in eating habits and inspire positive eating behaviors.

## 7.3 Human values proposition can inform play in designing interactive eating experiences

Human values refer to "desirable, trans-situational goals, serving as standards guiding people's actions". These values are abstract ideals that have the capacity to align people's emotions and attitudes while guiding behaviors that endure over time [31]. Consistent with the understanding that human values are socially desirable, but implicit, they can be seen as concepts that are used to drive behaviors both at the individuals and societal levels [30, 31]. Human values have been increasingly considered in the design of technology within the HCI community, with research focusing on the values of privacy [1, 35], universal usability [66], ethics [26] and trust [61]. In response to this increasing interest, Friedman et al. [25] proposed the term "value sensitive design". Building on this work, Mahamuni et al. [46] proposed that researchers and designers should consider the incorporation of human values when developing interactive systems, demonstrating that human values can elucidate and explain how people interact with digital technology [66]. Inspired by these works, we advocate for the inclusion of human values in the design of interactive eating and drinking systems in HFI. In particular, our research findings underline this perspective, as they suggest that integrating human values into the design of such systems can lead to a meaningful experience. For example, participants expressed enjoyment and excitement when using our system, reporting a sense of novelty and stimulation in their dining experiences. This positive response is indicative of an alignment with values related to pleasure and indulgence, as participants were able to savor a variety of foods and beverages in a self-indulgent manner during the study. Furthermore, our participants demonstrated a heightened ability to explore their eating activities creatively and spontaneously while engaging with the system. This creative engagement resonates with the value of self-expression and reflexivity, as individuals were given the freedom to personalize their dining experiences. Our findings suggest that aligning these systems with human values, such as pleasure, indulgence, and self-expression, can lead to more engaging and enjoyable eating and drinking experiences. As the field of HFI continues to advance, a deeper understanding of the interplay between human values and interactive eating systems might contribute to the development of human-food technologies that facilitate pleasurable dining experiences.

*7.3.1 Design implication: consider human values underlying gustosonic experiences.* Considering human values when designing interactive eating technology offers a new perspective for designers and researchers, particularly, when integrating machine learning models into newly designed eating technologies. With such advanced technology, prior research from the HFI community advocated that designing interactive eating technology should focus more on personalized, pleasurable and meaningful experiences in everyday life [4, 5]. We believe that designing an intelligent eating system presents a novel opportunity to deeply align with human values, enhancing the dining experience. By leveraging machine learning or generative AI technology, designers could create systems that adapt to individual preferences and cultural norms, fostering personalization and possibly a sense of connection. This approach allows for the dynamic integration of values such as pleasure, indulgence, and social interaction, making eating/drinking more than just a physical necessity but an enriching and pleasurable experience. Future designs could focus on how machine learning can interpret and respond to subtle user behaviors and preferences, transforming everyday eating into a meaningful, value-driven activity. This direction not only enriches the dining experience but could also set a new standard for how interactive eating technology aligns with our basic human needs.

## 8 LIMITATIONS AND FUTURE WORK

While we acknowledge that our work is not a comprehensive investigation into designing an interactive dining system, we contend that it can serve as a springboard for further HFI explorations. In particular, we acknowledge that our work focused on an individual's eating activities, at least in part due to the use of earbuds without audio pass-through functionality, which can isolate the individual from their environment. Future work could explore more social eating interactions, including with multiple participants, for example by using pass-through audio technology. We also acknowledge that, although our participants appreciated the use of the system across multiple meals, longer usage examination, possibly with additional recognition activities and sounds, could add further insights to our understanding of the design of interactive dining systems. Although many of our participants came originally from different cultural backgrounds, future studies might benefit from investigating such interactive dining systems in different cultural settings, because we know that values are highly culturally embedded. These cultural differences are important to consider also in regard to the use of sounds, because sounds interpretation is based on a cultural understanding. Furthermore, such future studies might also reveal insights from people who are less technology-experienced than our participant group. Our observations and discussions suggested to us that our participants were very familiar with technology, especially considering their age, and certainly not technology averse. We therefore encourage conducting longer-term studies in different geographical locations and with differently diverse participants.

We used a machine learning model to support eating any type of food and any type of drinks and accommodate dietary requirements and allergies, in contrast to prior HFI work, which required that participants eat specific food items such as ice cream and cookies (e.g. [55, 85, 86]. We demonstrated that our particular machine learning implementation, although certainly not perfect, is already sufficient to contribute to a playful experience. Future work could investigate advanced algorithms and models, possibly further contributing to the resulting user experience. For example, the random forest algorithm used in our backend server for the prediction of activities achieved an F1 score of 0.86, hence our system can report false positives and false negatives. While our participants told us that they did not see these errors as bugs, and that they facilitated serendipity and contributed to "stimulation", future studies might investigate how people perceive systems with more (or even less) accuracy. Furthermore, we acknowledge that our 10-fold cross-validation was a compromise, because we examined the larger sample although we are aware that it might yield different results in a different sample size. In this regard, future work could introduce and collect data on more food textures and also try different implementation and validation strategies in order to examine the results in terms of altered user experience. Furthermore, machine learning researchers could use our work as starting point to investigate more refined and elaborate ways to detect and classify eating and drinking actions. Future work could also investigate a different form factor for the earbuds, given that four of our participants found them uncomfortable to wear over longer periods. This issue speaks to prior work that highlighted that the form factor of wearables [65] is often overlooked, and extends the scope of considerations to include to the use of wearables as part of interactive dining systems.

We acknowledge that playful eating has the potential to encourage unhealthy eating behaviors, where treating eating as a reward could promote overconsumption or preference for nutritionally poor food. This concern is heightened by the increasing global rates of obesity and eating disorders [38]. While the technology offers potential benefits, future work could aim to incorporate a more intelligent sound feedback mechanism that encourages balanced and healthy eating patterns through playful design and machine learning. According to sensory expectation theory [69], sound can influence our judgments about food [70] and can evoke specific emotions about our food experiences [71]. Previous research has also demonstrated that sound and music can be effective in treating eating disorders [12, 59]. The future work of interactive eating technology could enable individuals to have control over the system and set their sound preferences based on dietary needs. Therefore, further research is required to explore the full range of effects that playful eating technology may have on eating behaviors. As such, we encourage others to collaborate in investigating how playful design can contribute to the creation of technologies that engage human sensory experiences and support overall wellbeing.

## 9 CONCLUSION

This work contributes to our understanding of the design of interactive dining systems, facilitated by a novel system called GustosonicSense. GustosonicSense draws from "gustosonic interactions", with sound based on eating and drinking actions delivered via earbuds, to facilitate playful experiences. Our study revealed how specific features of our design appeared to facilitate a playful eating and drinking experience, and these features and aspects of the experience were explained through "stimulation", "hedonism", and "reflexivity". We further emphasize the need for playful eating experiences and aim to inspire future design applications. For example, we can envision our system being used to help with dietary goals by playing motivational sound clips when eating healthy food, and playing discouraging sound clips when eating junk food. We can also envision our system being helpful in sensory learning for young children [36], for example, by incorporating auditory feedback related to taste, it could make connections between the food they eat, their cultural background, and the science behind food cultivation. We can also envision our system being helpful in aiding individuals or patients with impaired taste or smell [73], or those recovering from illness that affect the senses, by providing auditory cues that enhance the enjoyment of food. We can also envision our system being helpful in enriching the sensory experiences of astronauts by offering a variety of auditory stimuli while they consume the limited food options in space [58], potentially improving their wellbeing during long missions. Furthermore, we can envision social versions of our system where one person gets to hear whenever their remote partner eats, facilitating novel ways of social connections over a distance [50]. We believe that the results of our study will help designers and researchers to better understand augmented playful multisensory experiences around eating and drinking. These results could also be useful for interaction design and play researchers interesting in understanding how to facilitate playful experiences through technology augmentation of everyday life practices.

## ACKNOWLEDGMENTS

Humphrey O Obie was supported by ARC Discovery Project DP2001 00020. Zhuying Li is supported by the National Natural Science Foundation of China (Grant No.62302094). John Grundy is supported by ARC Laureate Fellowship FL190100035. Flora Salim acknowledges ARC CoE ADM+S (CE200100005) and the Nokia Bell Labs for the earables device donation. Florian 'Floyd' Mueller thanks the Australian Research Council, especially DP190102068, DP2001 02612 and LP210200656. We would like to thank all the participants.

## REFERENCES

[1] Mark S Ackerman and Lorrie Cranor. 1999. Privacy critics: UI components to safeguard users' privacy. In *CHI'99 Extended Abstracts on Human Factors in Computing Systems*. 258–259.

[2] Ferran Altarriba Bertran, Jared Duval, Katherine Isbister, Danielle Wilde, Elena Márquez Segura, Oscar Garcia Pañella, and Laia Badal León. 2019. Chasing play potentials in food culture to inspire technology design. In *Extended Abstracts of the Annual Symposium on Computer-Human Interaction in Play Companion Extended Abstracts*. 829–834.

[3] Ferran Altarriba Bertran, Sanvid Jhaveri, Rosa Lutz, Katherine Isbister, and Danielle Wilde. 2019. Making sense of human-food interaction. In *Proceedings of the 2019 CHI Conference on Human Factors in Computing Systems*. 1–13.

[4] Ferran Altarriba Bertran, Elena Márquez Segura, and Katherine Isbister. 2020. Technology for situated and emergent play: A bridging concept and design agenda. In *Proceedings of the 2020 CHI Conference on Human Factors in Computing Systems*. 1–14.

[5] Ferran Altarriba Bertran, Danielle Wilde, Ernő Berezvay, and Katherine Isbister. 2019. Playful human-food interaction research: State of the art and future directions. In *Proceedings of the Annual Symposium on Computer-Human Interaction in Play*. 225–237.

[6] Peter Arnold, Rohit Ashok Khot, and Florian 'Floyd' Mueller. 2018. "You Better Eat to Survive" Exploring Cooperative Eating in Virtual Reality Games. In *Proceedings of the Twelfth International Conference on Tangible, Embedded, and Embodied Interaction*. 398–408.

[7] Juha Arrasvuori, Marion Boberg, Jussi Holopainen, Hannu Korhonen, Andrés Lucero, and Markus Montola. 2011. Applying the PLEX framework in designing for playfulness. In *Proceedings of the 2011 Conference on Designing Pleasurable Products and Interfaces*. 1–8.

[8] Audacity. 2021. Free,open source, cross-platform audio software. https://www.audacityteam.org/.

[9] Jean-François Augoyard. 2014. *Sonic experience: a guide to everyday sounds*. McGill-Queen's Press-MQUP.

[10] Maribeth Back and D Des. 1996. *Micro-narratives in sound design: Context, character, and caricature in waveform manipulation*. Georgia Institute of Technology.

[11] Emilie Baltz. 2021. LICKESTRA – Emilie Baltz. http://emiliebaltz.com/experiments/lickestra/.

[12] Susanne Bauer. 2010. Music Therapy and Eating Disorders-A Single Case Study about the Sound of Human Needs. In *Voices: a world forum for music therapy*, Vol. 10.

[13] Steve Benford, Richard Ramchurn, Joe Marshall, Max L Wilson, Matthew Pike, Sarah Martindale, Adrian Hazzard, Chris Greenhalgh, Maria Kallionpää, Paul Tennent, et al. 2020. Contesting control: Journeys through surrender, self-awareness and looseness of control in embodied interaction. *Human–Computer Interaction* (2020), 1–29.

[14] Ferran Altarriba Bertran and Danielle Wilde. 2018. Playing with Food: reconfiguring the gastronomic experience through play. In *Proceedings of the 1st International Conference on Food Design and Food Studies (EFOOD 2017), October*.

[15] Shengjie Bi, Tao Wang, Nicole Tobias, Josephine Nordrum, Shang Wang, George Halvorsen, Sougata Sen, Ronald Peterson, Kofi Odame, Kelly Caine, et al. 2018. Auracle: Detecting eating episodes with an ear-mounted sensor. *Proceedings of the ACM on Interactive, Mobile, Wearable and Ubiquitous Technologies* 2, 3 (2018), 1–27.

[16] Patrick Biernacki and Dan Waldorf. 1981. Snowball sampling: Problems and techniques of chain referral sampling. *Sociological Methods & Research* 10, 2 (1981), 141–163.

[17] Dieuwerke P Bolhuis and Ciarán G Forde. 2020. Application of food texture to moderate oral processing behaviors and energy intake. *Trends in Food Science & Technology* 106 (2020), 445–456.

[18] J A Brillat-Savarin. 1835. Physiologie du goût [The philosopher in the kitchen/The physiology of taste]. *JP Meline: Bruxelles. Translated by A. Lalauze (1884), A handbook of gastronomy. Nimmo & Bain, London, UK* (1835).

[19] Yoram Chisik, Patricia Pons, and Javier Jaen. 2018. Gastronomy meets ludology: towards a definition of what it means to play with your (digital) food. In *Proceedings of the 2018 Annual Symposium on Computer-Human Interaction in Play Companion Extended Abstracts*. 155–168.

[20] Keum San Chun, Sarnab Bhattacharya, and Edison Thomaz. 2018. Detecting eating episodes by tracking jawbone movements with a non-contact wearable sensor. *Proceedings of the ACM on interactive, mobile, wearable and ubiquitous technologies* 2, 1 (2018), 1–21.

[21] Frédérique De Vignemont. 2010. Body schema and body image—Pros and cons. *Neuropsychologia* 48, 3 (2010), 669–680.

[22] Oliver Doehrmann and Marcus J Naumer. 2008. Semantics and the multisensory brain: how meaning modulates processes of audio-visual integration. *Brain research* 1242 (2008), 136–150.

[23] John Ferrara. 2012. *Playful design: Creating game experiences in everyday interfaces*. Rosenfeld Media.

[24] Robin Fox. 2003. Food and eating: an anthropological perspective. *Social Issues Research Centre* 2003 (2003), 1–21.

[25] Batya Friedman, Peter Kahn, and Alan Borning. 2002. Value sensitive design: Theory and methods. *University of Washington technical report* 2 (2002), 12.

[26] Batya Friedman and Peter H Kahn Jr. 2003. Human values, ethics, and design. *The human-computer interaction handbook* (2003), 1177–1201.

[27] Andrea Grimes and Richard Harper. 2008. Celebratory technology: new directions for food research in HCI. In *Proceedings of the SIGCHI conference on human factors in computing systems*. 467–476.

[28] Greg Guest, Kathleen M MacQueen, and Emily E Namey. 2011. *Applied thematic analysis*. Sage publications.

[29] Lilit Hakobyan, Jo Lumsden, Rachel Shaw, and Dympna O'Sullivan. 2016. A longitudinal evaluation of the acceptability and impact of a diet diary app for older adults with age-related macular degeneration. In *Proceedings of the 18th international conference on human-computer interaction with mobile devices and services*. 124–134.

[30] Paul HP Hanel, Lukas F Litzellachner, and Gregory R Maio. 2018. An empirical comparison of human value models. *Frontiers in Psychology* (2018), 1643.

[31] Paul HP Hanel, Katia C Vione, Ulrike Hahn, and Gregory R Maio. 2017. Value instantiations: the missing link between values and behavior? In *Values and behavior*. Springer, 175–190.

[32] Yuki Hashimoto, Naohisa Nagaya, Minoru Kojima, Satoru Miyajima, Junichiro Ohtaki, Akio Yamamoto, Tomoyasu Mitani, and Masahiko Inami. 2006. Strawlike user interface: Virtual experience of the sensation of drinking using a straw. In *Proceedings of the 2006 ACM SIGCHI international conference on Advances in computer entertainment technology*. 50–es.

[33] Md. Shafiqul Islam. [n. d.]. Human Activity Recognition from eSense datasets. [Online.] Available: https://github.com/shafiqulislamsumon/HARESense. Accessed October, 2021.

[34] Shin-ichiro Iwatani, Hidemi Akimoto, and Naoki Sakurai. 2013. Acoustic vibration method for food texture evaluation using an accelerometer sensor. *Journal of food engineering* 115, 1 (2013), 26–32.

[35] Gavin Jancke, Gina Danielle Venolia, Jonathan Grudin, Jonathan J Cadiz, and Anoop Gupta. 2001. Linking public spaces: technical and social issues. In *Proceedings of the SIGCHI conference on Human factors in computing systems*. 530–537.

[36] Azusa Kadomura, Cheng-Yuan Li, Koji Tsukada, Hao-Hua Chu, and Itiro Siio. 2014. Persuasive technology to improve eating behavior using a sensor-embedded fork. In *Proceedings of the 2014 acm international joint conference on pervasive and ubiquitous computing*. 319–329.

[37] Fahim Kawsar, Chulhong Min, Akhil Mathur, Alessandro Montanari, Utku Günay Acer, and Marc Van den Broeck. 2018. esense: Open earable platform for human

sensing. In *Proceedings of the 16th ACM Conference on Embedded Networked Sensor Systems*. 371–372.

[38] Ronald C Kessler, Patricia A Berglund, Wai Tat Chiu, Anne C Deitz, James I Hudson, Victoria Shahly, Sergio Aguilar-Gaxiola, Jordi Alonso, Matthias C Angermeyer, Corina Benjet, et al. 2013. The prevalence and correlates of binge eating disorder in the World Health Organization World Mental Health Surveys. *Biological psychiatry* 73, 9 (2013), 904–914.

[39] Naoya Koizumi, Hidekazu Tanaka, Yuji Uema, and Masahiko Inami. 2011. Chewing Jockey. In *Proceedings of the 8th International Conference on Advances in Computer Entertainment Technology - ACE '11*. 1. https://doi.org/10.1145/2071423.2071449

[40] Stacey Kuznetsov, Christina J Santana, and Elenore Long. 2016. Everyday food science as a design space for community literacy and habitual sustainable practice. In *Proceedings of the 2016 CHI conference on human factors in computing systems*. 1786–1797.

[41] Hyosun Kwon, Shashank Jaiswal, Steve Benford, Sue Ann Seah, Peter Bennett, Boriana Koleva, and Holger Schnädelbach. 2015. FugaciousFilm: Exploring attentive interaction with ephemeral material. In *Proceedings of the 33rd Annual ACM Conference on Human Factors in Computing Systems*. 1285–1294.

[42] Zach Lieberman. 2021. Drink Up Fountain — YesYesNo Interactive Projects. http://www.yesyesno.com/drink-up-fountain.

[43] Robyn Longhurst. 2003. Semi-structured interviews and focus groups. *Key methods in geography* 3, 2 (2003), 143–156.

[44] Pedro Lopes, Josh Andres, Richard Byrne, Nathan Semertzidis, Zhuying Li, Jarrod Knibbe, Stefan Greuter, et al. 2021. Towards understanding the design of bodily integration. *International Journal of Human-Computer Studies* 152 (2021), 102643.

[45] Roya Lotfi, George Tzanetakis, Rasit Eskicioglu, and Pourang Irani. 2020. A comparison between audio and IMU data to detect chewing events based on an earable device. In *Proceedings of the 11th Augmented Human International Conference*. 1–8.

[46] Ravi Mahamuni, Kejul Kalyani, and Piyush Yadav. 2015. A simplified approach for making human values central to interaction design. *Procedia Manufacturing* 3 (2015), 874–881.

[47] Angelo Maravita, Charles Spence, and Jon Driver. 2003. Multisensory integration and the body schema: close to hand and within reach. *Current biology* 13, 13 (2003), R531–R539.

[48] Robb Mitchell, Alexandra Papadimitriou, Youran You, and Laurens Boer. 2015. Really eating together: A kinetic table to synchronise social dining experiences. In *Proceedings of the 6th Augmented Human International Conference*. 173–174.

[49] Florian'Floyd' Mueller, Marianna Obrist, Ferran Altarriba Bertran, Neharika Makam, Soh Kim, Christopher Dawes, Patrizia Marti, Maurizio Mancini, Eleonora Ceccaldi, Nandini Pasumarthy, et al. 2023. Grand Challenges in Human-Food Interaction. *International Journal of Human-Computer Studies* (2023).

[50] Florian Mueller, Frank Vetere, Martin R Gibbs, Darren Edge, Stefan Agamanolis, and Jennifer G Sheridan. 2010. Jogging over a distance between Europe and Australia. In *Proceedings of the 23nd annual ACM symposium on User interface software and technology*. 189–198.

[51] Florian 'Floyd' Mueller, Tuomas Kari, Zhuying Li, Yan Wang, Yash Dhanpal Mehta, Josh Andres, Jonathan Marquez, and Rakesh Patibanda. 2020. Towards Designing Bodily Integrated Play. In *Proceedings of the Fourteenth International Conference on Tangible, Embedded, and Embodied Interaction*. 207–218.

[52] Florian 'Floyd' Mueller, Yan Wang, Zhuying Li, Tuomas Kari, Peter Arnold, Yash Dhanpal Mehta, Jonathan Marquez, and Rohit Ashok Khot. 2020. Towards experiencing eating as play. In *Proceedings of the Fourteenth International Conference on Tangible, Embedded, and Embodied Interaction*. 239–253.

[53] Florian 'Floyd' Mueller, Nathan Semertzidis, Josh Andres, Joe Marshall, Steve Benford, Xiang Li, Louise Matjeka, and Yash Mehta. 2023. Toward Understanding the Design of Intertwined Human-Computer Integrations. *ACM Transactions on Computer-Human Interaction* 30, 5 (2023), 1–45.

[54] Martin Murer, Ilhan Aslan, and Manfred Tscheligi. 2013. LOLL io. In *Proceedings of the 7th International Conference on Tangible, Embedded and Embodied Interaction - TEI '13*. 299. https://doi.org/10.1145/2460625.2460675

[55] Takuji Narumi, Shinya Nishizaka, Takashi Kajinami, Tomohiro Tanikawa, and Michitaka Hirose. 2011. Meta cookie+: an illusion-based gustatory display. In *International Conference on Virtual and Mixed Reality*. Springer, 260–269.

[56] Jon Noronha, Eric Hysen, Haoqi Zhang, and Krzysztof Z Gajos. 2011. Platemate: Crowdsourcing Nutritional Analysis from Food Photographs. In *Proceedings of the 24th Annual ACM Symposium on User Interface Software and Technology*. 1–12. https://doi.org/10.1145/2047196.2047198

[57] NVivo. 2021. Qualitative Data Analysis Software | NVivo - QSR International. https://www.qsrinternational.com/nvivo-qualitative-data-analysis-software/home.

[58] Marianna Obrist, Yunwen Tu, Lining Yao, and Carlos Velasco. 2019. Space food experiences: designing passenger's eating experiences for future space travel scenarios. *Frontiers in Computer Science* 1 (2019), 3.

[59] Varvara Pasiali, Dean Quick, Jessica Hassall, and Hailey A Park. 2020. Music therapy programming for persons with eating disorders: a review with clinical examples. In *Voices: A World Forum for Music Therapy*, Vol. 20. 15–15.

[60] Pietro Polotti, Stefano Delle Monache, Stefano Papetti, and Davide Rocchesso. 2008. Gamelunch: Forging a dining experience through sound. In *CHI'08 extended abstracts on Human factors in computing systems*. 2281–2286.

[61] Elena Rocco. 1998. Trust breaks down in electronic contexts but can be repaired by some initial face-to-face contact. In *Proceedings of the SIGCHI conference on Human factors in computing systems*. 496–502.

[62] Yvonne Rogers, Nicola Yuill, and Paul Marshall. 2013. Contrasting lab-based and in-the-wild studies for evaluating multi-user technologies. *The SAGE handbook of digital technology research, SAGE, London* (2013), 359–373.

[63] Gery W Ryan and H Russell Bernard. 2003. Techniques to identify themes. *Field methods* 15, 1 (2003), 85–109.

[64] Bert Schiettecatte and Jean Vanderdonckt. 2008. AudioCubes: A distributed cube tangible interface based on interaction range for sound design. In *Proceedings of the 2nd international conference on Tangible and embedded interaction*. 3–10.

[65] Nathan Arthur Semertzidis, Annaelle Li Pin Hiung, Michaela Jayne Vranic-Peters, and Florian 'Floyd' Mueller. 2023. Dozer: Towards understanding the design of closed-loop wearables for sleep. In *Proceedings of the 2023 CHI Conference on Human Factors in Computing Systems*. 1–14.

[66] Ben Shneiderman. 2000. Universal usability. *Commun. ACM* 43, 5 (2000), 84–91.

[67] Charles Spence. 2010. The multisensory perception of flavour. *Psychologist* 23, 9 (2010).

[68] Charles Spence. 2011. Crossmodal correspondences: A tutorial review. *Attention, Perception, & Psychophysics* 73, 4 (2011), 971–995.

[69] Charles Spence. 2012. Managing sensory expectations concerning products and brands: Capitalizing on the potential of sound and shape symbolism. *Journal of Consumer Psychology* 22, 1 (2012), 37–54.

[70] Charles Spence. 2015. Eating with our ears: assessing the importance of the sounds of consumption on the perception and enjoyment of multisensory flavour experiences. *Flavour* 4, 1 (2015), 3.

[71] Charles Spence. 2015. Sound Bites & Digital Seasoning. *Sound and Interactivity* 20 (2015), 9.

[72] Charles Spence. 2016. Sound: The forgotten flavor sense. In *Multisensory Flavor Perception*. Elsevier, 81–105.

[73] Charles Spence. 2017. Comfort food: A review. *International journal of gastronomy and food science* 9 (2017), 105–109.

[74] Charles Spence. 2017. *Gastrophysics: The new science of eating*. Penguin UK.

[75] Charles Spence. 2021. Sonic Seasoning and Other Multisensory Influences on the Coffee Drinking Experience. *Frontiers in Computer Science* 3 (2021), 21.

[76] Charles Spence, Maurizio Mancini, and Gijs Huisman. 2019. Digital commensality: Eating and drinking in the company of technology. *Frontiers in psychology* 10 (2019), 2252.

[77] Charles Spence and Betina Piqueras-Fiszman. 2014. *The perfect meal: the multisensory science of food and dining*. John Wiley & Sons.

[78] Charles Spence, Felipe Reinoso-Carvalho, Carlos Velasco, and Qian Janice Wang. 2019. Extrinsic auditory contributions to food perception & consumer behaviour: An interdisciplinary review. *Multisensory research* 32, 4-5 (2019), 275–318.

[79] Chie Suzuki, Takuji Narumi, Tomohiro Tanikawa, and Michitaka Hirose. 2014. Affecting Tumbler: Affecting our flavor perception with thermaldback. In *Proceedings of the 11th Conference on Advances in Computer Entertainment Technology*. 1–10.

[80] Thomas Tannou, Thomas Lihoreau, Mélanie Couture, Sylvain Giroux, Guillaume Spalla, Sareh Zarshenas, Mireille Gagnon-Roy, Aline Aboujaoudé, Amel Yaddaden, Lucas MORIN, et al. 2022. Is research on 'smart living environments' based on unobtrusive technologies for older adults going in circles? Evidence from an umbrella review. *Ageing Research Reviews* (2022), 101830.

[81] Andries van der Bilt and JH Abbink. 2017. The influence of food consistency on chewing rate and muscular work. *Archives of oral biology* 83 (2017), 105–110.

[82] Shawn VanCour and Kyle Barnett. 2017. Eat what you hear: Gustasonic discourses and the material culture of commercial sound recording. *Journal of Material Culture* 22, 1 (2017), 93–109. https://doi.org/10.1177/1359183516679186

[83] Carlos Velasco, Felipe Reinoso Carvalho, Olivia Petit, and Anton Nijholt. 2016. A Multisensory Approach for the Design of Food and Drink Enhancing Sonic Systems. *Proceedings of the 1st Workshop on Multi-sensorial Approaches to Human-Food Interaction MHFI '16* (2016), 1–7. https://doi.org/10.1145/3007577.3007578

[84] Carlos Velasco, Anton Nijholt, Charles Spence, Takuji Narumi, Kosuke Motoki, Gijs Huisman, and Marianna Obrist. 2020. Multisensory approaches to human-food interaction. In *Proceedings of the 2020 International Conference on Multimodal Interaction*. 878–880.

[85] Yan Wang, Zhuying Li, Robert S Jarvis, Joseph La Delfa, Rohit Ashok Khot, and Florian 'Floyd' Mueller. 2020. WeScream! Toward Understanding the Design of Playful Social Gustosonic Experiences with Ice Cream. In *Proceedings of the 2020 ACM Designing Interactive Systems Conference*. 951–963.

[86] Yan Wang, Zhuying Li, Robert S Jarvis, Angelina Russo, Rohit Ashok Khot, and Florian 'Floyd' Mueller. 2019. Towards Understanding the Design of Playful Gustosonic Experiences with Ice Cream. In *Proceedings of the Annual Symposium on Computer-Human Interaction in Play*. 239–251.

[87] Yan Wang, Zhuying Li, Rohit Ashok Khot, and Florian'Floyd' Mueller. 2022. Toward Understanding Playful Beverage-based Gustosonic Experiences. *Proceedings of the ACM on Interactive, Mobile, Wearable and Ubiquitous Technologies* 6, 1 (2022), 1–23.

[88] Yan Wang, Xian Zhang, Zhuying Li, Rohit Ashok Khot, and Florian 'Floyd' Mueller. 2020. Towards a Framework for Designing Playful Gustosonic Experiences. In *Extended Abstracts of the 2020 CHI Conference on Human Factors in Computing Systems*. 1–9.

[89] Danielle Wilde and Ferran Altarriba Bertran. 2019. Participatory Research through Gastronomy Design: a designerly move towards more playful gastronomy. *International Journal of Food Design* 4, 1 (2019), 3–37.

[90] Guoqiang Zhang, Michael Y Hu, B Eddy Patuwo, and Daniel C Indro. 1999. Artificial neural networks in bankruptcy prediction: General framework and cross-validation analysis. *European journal of operational research* 116, 1 (1999), 16–32.

[91] Rui Zhang, Severin Bernhart, and Oliver Amft. 2016. Diet eyeglasses: Recognising food chewing using EMG and smart eyeglasses. In *2016 IEEE 13th International Conference on Wearable and Implantable Body Sensor Networks (BSN)*. IEEE, 7–12.

[92] Shibo Zhang, Yuqi Zhao, Dzung Tri Nguyen, Runsheng Xu, Sougata Sen, Josiah Hester, and Nabil Alshurafa. 2020. Necksense: A multi-sensor necklace for detecting eating activities in free-living conditions. *Proceedings of the ACM on interactive, mobile, wearable and ubiquitous technologies* 4, 2 (2020), 1–26.

[93] Ya-Li Zheng, Xiao-Rong Ding, Carmen Chung Yan Poon, Benny Ping Lai Lo, Heye Zhang, Xiao-Lin Zhou, Guang-Zhong Yang, Ni Zhao, and Yuan-Ting Zhang. 2014. Unobtrusive sensing and wearable devices for health informatics. *IEEE transactions on biomedical engineering* 61, 5 (2014), 1538–1554.